\documentclass[useAMS,usenatbib]{mn2e}
\usepackage{psfig, epsf, epsfig}
\title[Dark impact]{Dark impact and galactic star formation:
Origin of the Gould belt}
\author[K. Bekki
]{Kenji Bekki${}^1$\thanks{E-mail:
bekki@phys.unsw.edu.au} \\
       ${}^1$School of Physics, University of New South Wales,
              Sydney 2052, NSW, Australia}

\begin{document}

\date{Accepted, Received 2005 February 20; in original form }

\pagerange{\pageref{firstpage}--\pageref{lastpage}} \pubyear{2005}

\maketitle
\label{firstpage}
\begin{abstract}

The Milky Way has a giant stellar structure  in the solar neighborhood,
which has
a size of $\sim 1$ kpc, a mass of $\sim 10^6 {\rm M}_{\odot}$,
and a ring-like distribution of  young stars.
Fundamental physical properties
of this  local enigmatic structure,  known as the Gould belt (GB),
have  not been reproduced by previously proposed models. 
We first  show that the 
local enigmatic structure
can be  formed about 30 Myr ago
as a result of a high-speed, oblique  collision between
a gas cloud with a mass of $\sim 10^6 {\rm M}_{\odot}$
and a dark matter clump with a mass of $\sim  10^7 {\rm M}_{\odot}$
based on numerical
simulations of the collision.  We find that strong dynamical impact
of the clump transforms the flattened cloud into a ring-like stellar
structure
after induced star formation within the cloud.
Our simulations furthermore demonstrate that
the stellar structure is moderately
elongated and
significantly inclined  with respect to the disk of the Milky Way
owing to the strong tidal torque by the colliding  clump.
We thus suggest that the GB is one of  stellar
substructures
formed  from collisions between gas clouds and dark
matter clumps predicted in the hierarchical clustering  scenario
of  galaxy formation.
We also suggest that collisions of dark matter clumps with their host
galaxies
can significantly change star formation histories
for some of  their gas clouds thus influence  galactic global  star
formation histories
to some extent.
Our simulations  predict that unique  giant stellar substructures
recently discovered in other galaxies can result from
dynamical impact of their dark matter clumps on their gas clouds.

\end{abstract}


\section{Introduction}

The GB is  a flattened stellar structure with
an inclination of $16^{\circ}-22^{\circ}$
to the Galactic plane
and contains about 60\% of young  stars with ages of $30-60$ Myr
within 600 pc from the Sun (e.g., Olano 1982; Poppel 1997;
Guillout et al. 1998;
Torra et al. 2000).
The projected distribution of young  stars in the GB  is suggested to be
quite inhomogeneous and  significantly
elongated with an ellipticity of $\sim 0.3$ (e.g., Poppel 1997).
The observed  stellar kinematic of the GB suggests that
the GB has internal rotation with a possible angular velocity
of $7-24$ km s$^{-1}$ kpc$^{-1}$
and is currently expanding (Lindblad 2000).
The GB is considered to be young with its age less than
$30-60$ Myr (e.g., Poppel 1997).
Previous theoretical studies suggested that the formation of the GB
is closely associated  with
efficient star  formation in an expanding gaseous ring
formed as a result of stellar feedback effects of
massive OB stars and supernovae on interstellar medium
in the solar neighborhood (e.g., Poppel 1997).
Other studies (e.g., Come\'on \&  Torra 1994)
proposed an oblique impact  of a high-velocity cloud on the gas in the Galactic
disk for the GB formation.

\begin{figure*}
\psfig{file=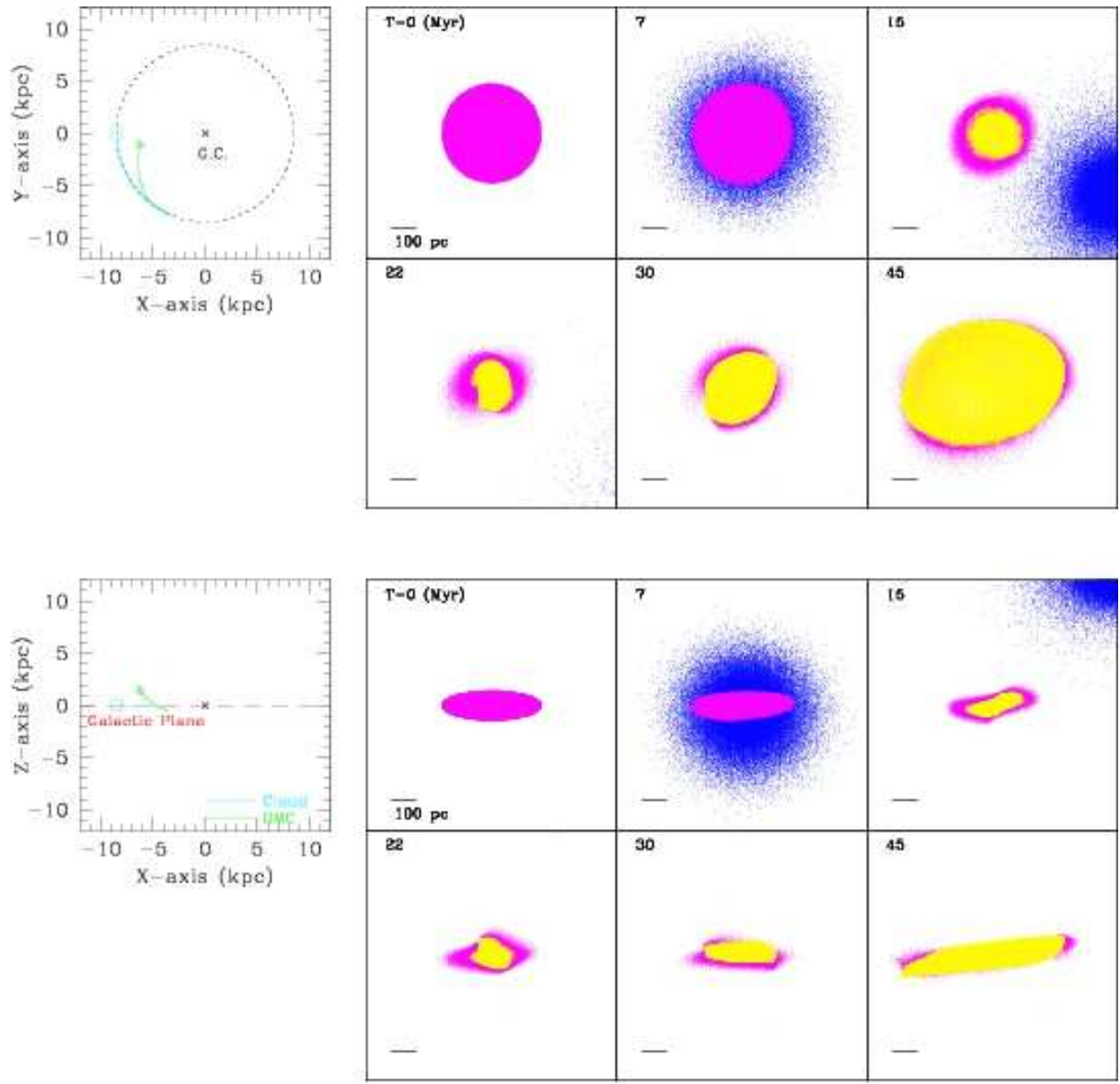,width=16.0cm}
\caption{
Morphological evolution of a gas cloud colliding with
a dark matter clump (DMC) projected onto the $X$-$Y$ plane (upper)
and the $X$-$Z$ one (lower).
The time $T$ (in Myr) shown in the upper left corner of each panel
represents the time that has elapsed since the simulation
started.
The two separate panels show the orbits
of the cloud (cyan) and the DMC (green) projected onto
the $X$-$Y$ and  the $X$-$Z$ planes.
The final locations
of the cloud and the DMC indicated by an open square and
and an open triangle, respectively,
and the solar radius
($R=8.5$ kpc corresponding to the distance between the Sun and the
Galactic center, G.C.) and the Galactic plane are shown in these
two panels for comparison.
The cloud and the DMC are represented
initially by 150000 gaseous particles and 100000 stellar ones,
respectively,
and shown by magenta and blue, respectively. The yellow particles are
new stars formed from gas  in the cloud.
A bar shown in the lower left corner of each frame measures
100 pc.
The time evolution of the cloud-DMC distance is described  in Fig. 3,
which clearly shows that the cloud-DMC collision is an off-center one.
}
\label{Figure. 1}
\end{figure*}

\begin{figure*}
\psfig{file=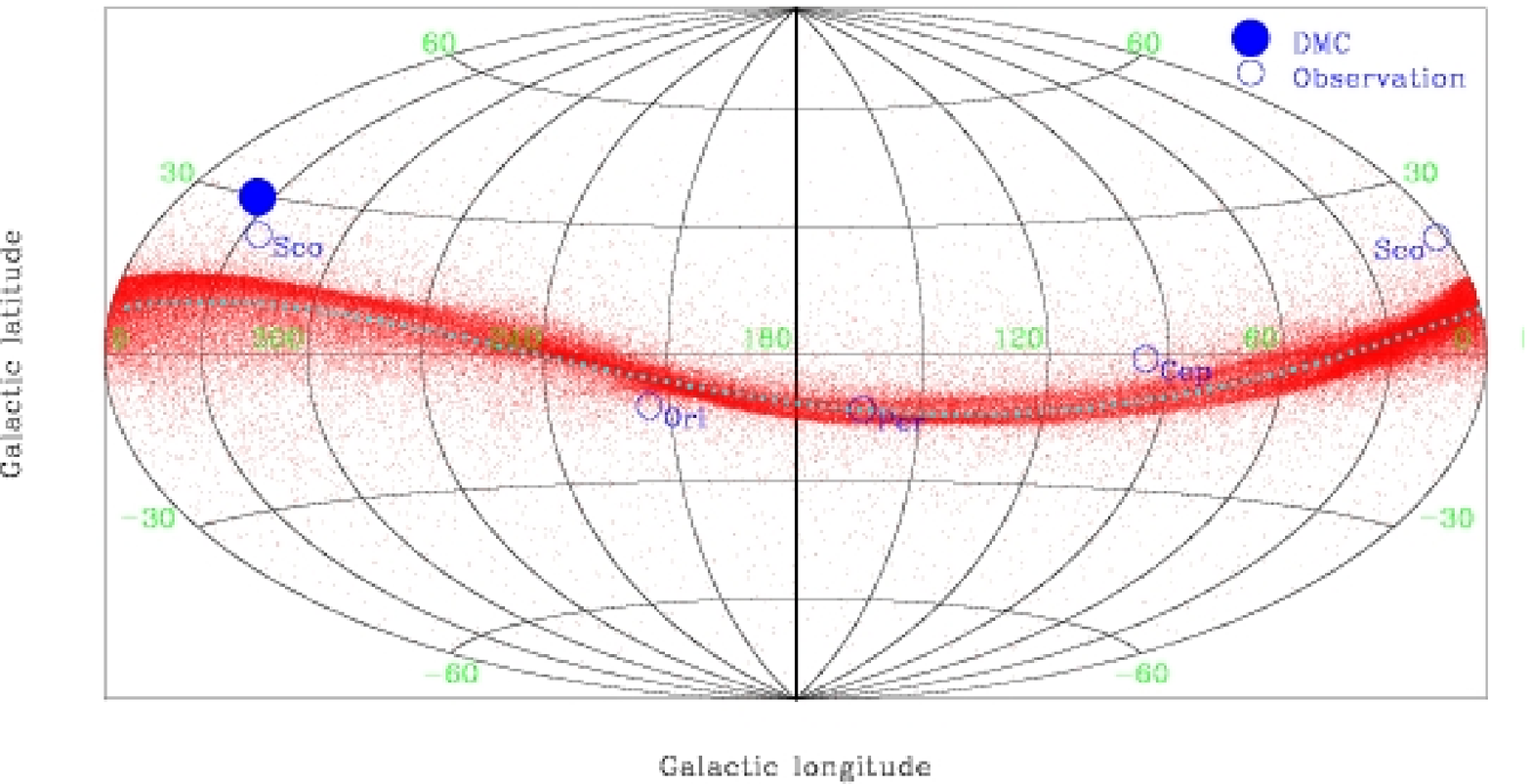,width=15.0cm}
\caption{
 Final distribution of stellar particles in
the cloud at $T=45$ Myr
in the Galactic coordinate system, where abscissa and ordinate denote
Galactic longitude  ($l$) and latitude ($b$), respectively.
All stellar and gaseous particles
are shown by red and the final location
of the  DMC is shown by a filled blue circle for
comparison. The five blue open circles represent the observed
OB associations on the Gould Belt (de Zeeuw et al. 1999): two OB2 in Scorpius
constellation (Sco), OB 1 in Orion (Ori), OB2 in Perseus (Per),
and OB2 in Cepheus (Cep). 
Only one of the listed locations 
in de Zeeuw et al. (1999) for each OB association
is shown here for convenience
(e.g., ($l$,$b$)=($164^{\circ}$,$-13^{\circ}$) for Per OB2).
The dotted thick cyan line describes
the best-fit Gould curve (Taylor et al. 1987)  of $b=b_0 \sin ( \frac{\pi}{180}
[l-l_{0}]  ) +
b_{\rm c} $
for the simulated stellar belt,
where $b_0$ ($\approx 12^{\circ}$), $l_0$ ($\approx 232^{\circ}$),
and $b_{\rm c}$  ($\approx -2^{\circ}$)
denote the inclination angle of the belt with respect to the Galactic
plane,
the ascending node longitude of the belt,
and the vertical distance between the mass center of the belt
and the Galactic plane,
respectively.
The simulated Gould curve is more similar to that
derived from  the observed molecular clouds (Taylor et al. 1987)
than to  those from young stars (Poppel 1997; Perrot \& Grenier 2003).
}
\label{Figure. 2}
\end{figure*}

The purpose of this paper is thus
to propose a new scenario which explains self-consistently
the observed fundamental properties of the GB.
In the new scenario presented in this paper,
the GB is one of  stellar substructures (e.g., rings and arcs)
formed from high-speed, off-center collisions between giant molecular
clouds, which are believed to be the birth places for stars (Lada \&
Lada 2003),
and dark matter clumps (DMCs) orbiting the Galaxy.
This dynamical impact of DMCs is referred to as the dark impact
for convenience in Bekki \& Chiba (2006)
and used as such in  the present paper.
Since recent high-resolution cosmological simulations of the Galactic
halo formation
based on the cold dark matter model (CDM)
have shown that
hundreds of high-density
DMCs may exist in the solar-neighborhood (Diemand et al. 2008),
we consider that such collisions between gas clouds and DMCs are
almost inevitable.
We here do not intend to discuss the well known over-abundance problem
of satellite galaxies (hosted by dark matter subhalos)
in the CDM,  just because this is beyond the scope of this paper.

\section{The model}

We here show the results of GRAPE-SPH simulations
(Bekki \& Chiba 2007) only  for the  best model explaining
the observed properties of the GB self-consistently,
and will describe those of other models with different
model parameters in our future study (e.g., Bekki 2009 in preparation,
B09).
The progenitor cloud of the GB is modeled as an oblate
sphere with a radius of 200 pc,
an ellipticity of 0.3, a mass of $10^6 {\rm M}_{\odot}$,
temperature of 100K,
and rigid rotation with an  angular velocity of $4.8 \times 10^{-2}$
km s$^{-1}$ pc$^{-1}$.
These parameters values are consistent with the observed properties
of GMCs (e.g., Solomon et al. 1979;
Kerton et al. 2003;  Rosolowsky et al. 2003).
GMCs with sizes of $\sim$  100pc exist 
in the Galaxy (e.g., Fig. 6 in Solomon et al. 1979),
and those with sizes of $100-200$pc exist in other galaxies
(e.g., NGC 4449 and NGC4605; see Fig. 1 in Bolatto et al. 2008).
Therefore, given the observed uncertainties in estimating
the GMC sizes (Bolatto et al. 2008),
the adopted cloud sizes are consistent 
with observations for very large GMCs.
Star formation is assumed to happen in local regions
of the cloud if the regions satisfy physical conditions
for the Jeans instability that can cause collapse of
interstellar gas and subsequent star formation (e.g., Bekki \& Chiba
2007).

The self-gravitating
cloud is dynamically influenced by the fixed gravitational potential
of the Galaxy composed of bulge, disk, and dark matter halo.
We adopt the same Galactic potential as that used 
in Helmi \& de Zeeuw (2000) which investigated orbits
of dwarfs galaxies in the Galactic halo.
The cloud is located initially at 8.5 kpc
from the Galactic center (G.C.) and moves round  the Galaxy with
a circular speed of 220 km s$^{-1}$.
The present location of the Sun is set to be ($X$, $Y$, $Z$)=($-8.5$, 0,
0) kpc
and the Galaxy rotates to the +Y direction in the present study.
The initial location 
and velocity for the cloud
are chosen such that the final
(i.e., present) location can be  almost the same as that of
the Sun after its 45 Myr evolution (see Fig.1 for the initial
three-dimensional
position  in the Galaxy).

The DMC has a mass of $3 \times 10^7 {\rm
M}_{\odot}$, a mean density of $2.0 {\rm M}_{\odot}$ pc$^{-3}$
within 100pc,
and the NFW radial density profile (Navarro et al. 1996)
within its tidal radius ($=317$ pc), which are all consistent
with predictions from the latest cosmological simulations (Diemand et
al. 2008; Springel et
al. 2008a).
The initial relative position ($X_{\rm r}$,  $Y_{\rm r}$, $Z_{\rm r}$)
and velocity ($U_{\rm  r}$,  $V_{\rm r}$, $W_{\rm r}$) of the DMC with
respect to the cloud are set to be
($-446$, $233$, $-600$) pc and
($59$, $-29$, $66$) km s$^{-1}$, respectively.
For these relative positions and velocities,
the DMC collides with the cloud at $T \approx8$ Myr, where $T$
represents the time  that has elapsed since
the simulation started.

We have confirmed that (i) the ring-like structure can be formed
in models with a wider range of model parameters (e.g., impact
parameter) and (ii) the dark impact of DMCs with
masses of $\sim 10^7 {\rm M}_{\odot}$
can not influence structure and kinematics of the old
stellar disk  (B09).
The origin of 
unique  giant stellar substructures
like the GB found in disks for a number
of galaxies (Efremov\&  Elmegreen 1998;
Comer\'on 2001;  Larsen et al. 2002)
and the ``shingles'' in the Galaxy 
(e.g., Schmidt-Kaler \&  Schlosser 1973 )
can be also explained 
in the context of the dark impact (B09).
The simulated inclination angle of the GB is slightly smaller
than the observed one, which may be one of
disadvantages of the model: however,  the inclination angle
depends on (i) the time step at which it is estimated
and (ii) the assumed location of the Sun owing to the presession
of the GB.

\begin{figure}
\psfig{file=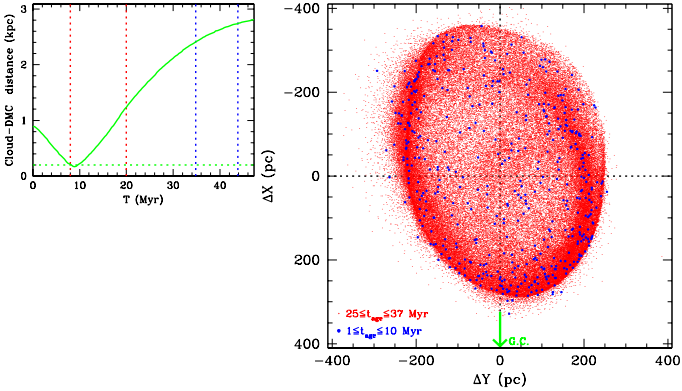,width=9.0cm}
\caption{
Time evolution of the distance between the gas cloud and the DMC
(left) and the  projected distribution of new stars in the simulated
stellar structure at $T=45$ Myr (right).
The dotted thick green line
indicates the initial cloud radius (=200 pc) and
the intersection between the line and the thick solid green line
for the cloud-DMC distance can indicate the time
when the cloud and the DMC start their strong dynamical interaction
in the left panel.
The dotted red lines indicate
the start and the end of
active star formation (i.e., starburst)
during the dark impact in the left panel.
The two dotted blue lines indicate a range of  younger
stellar populations with ages ($t_{\rm age}$) between
1 Myr and 10 Myr in the left panel.
The starburst population with 25 Myr $\le t_{\rm age}\le$ 37 Myr
and younger one with 1 Myr $\le t_{\rm age}\le$ 10 Myr are shown
by smaller red dots and bigger blue ones, respectively in
the right panel.
Here $\Delta X$ and $\Delta Y$ denote $X$- and  $Y$-positions
of particles with respect to the mass center of the stellar structure.
The direction of the Galactic center (G.C.) is
shown by a green arrow.
The center of the frame is set to be the mass center
of the stellar structure so that the simulated distribution can be
compared with the observed one (Poppel 1997; Perrot \& Grenier 2003).
}
\label{Figure. 3}
\end{figure}

\section{Results}

Fig. 1 shows that as the DMC passes through the cloud from its beneath
($T=7$ Myr), it first compresses the central region of the cloud
so that new stars can form from the high-density region of the cloud
($T=7-15$ Myr).
Star formation rate increases rapidly during the collision
($T=8$ Myr), reaches its maximum of 0.16 ${\rm M}_{\odot}$ yr$^{-1}$
($T=12$ Myr), and rapidly decrease
to be 0.005 ${\rm M}_{\odot}$ yr$^{-1}$
after the collision ($T=22$ Myr).
The dark  impact
then induces an  outwardly propagating density
wave and consequently transforms the oblate
gas cloud  into the  ring-like  one ($T=22$ Myr).
The tidal torque of the dark impact also causes the
precession of the internal spin vector of the new stellar system
so that the system can be seen as
significantly  inclined with respect to the Galactic plane ($T=22$ Myr).
The gas and new stars in the cloud  can
continue to expand after the dark impact,
while the inclination angle
of the stellar system becomes apparently smaller in
edge-on view owing to precession
($T=30$ Myr).
The strong tidal field of the  Galaxy
transforms the expanding stellar system into an elongated ring-like
structure
that is still inclined to the Galactic plane ($T$=45 Myr).

Fig. 2 shows that
the simulated ring-like structure can be seen
as  a giant arc across the Galactic plane
if it is projected onto the Galactic coordinate system.
The arc-like appearance is due to the disky  distribution of gas
and stars inclined to the Galactic plane.
The simulated arc-like belt can cover the observed OB stellar
associations
that are considered to be formed in the GB,
which suggests that the present model can reproduce well the
distribution of young stellar populations in the GB.
The DMC is currently  located at ($X$, $Y$, $Z$) = ($-6.3$, $-1.0$,
$1.4$) kpc
and the  distances of the DMC from the Galactic center and the Sun
are 6.5 and 2.8 kpc, respectively.
Fig. 3 shows that the projected distribution
of  older  stars  with stellar ages
($t_{\rm age}$) ranging from 25 Myr to 37 Myr (corresponding to
stars formed between $T$=8 Myr and 20 Myr, i.e., during a starburst)
with respect to the mass center of all stars
appears to be ringed and
moderately elongated.
The distribution of the very young stars with 1 Myr $\le t_{\rm age}
\le$ 10
Myr also appears to be ringed,
though a significant fraction of the stars exist inside the ring-like
structure.
The elongated morphology and size of the GB
in the present simulation
are  similar to those in  the pervious dynamical
models for the observed properties of the GB (Perrot \&
Grenier 2003).

Fig. 4 shows that almost all stars have positive radial velocities
($v_{\rm r}$) with a mean $v_{\rm r}$ of 7.1 km s$^{-1}$,
which means that the simulated stellar structure is expanding
with respect to its mass center.
The expansion rate of the structure at $\Delta Y=400$ pc
is 17.9  km s$^{-1}$ kpc$^{-1}$, which is quite similar to
that (20
km s$^{-1}$ kpc$^{-1}$) derived from one of  kinematical
models of the GB (Lindblad 2000).
Although there  is a weak tendency
that the outer stars are more likely to have larger $v_{\rm r}$,
the distributions of the stars in this $\Delta Y$-$v_{\rm r}$ phase
space
is highly inhomogeneous owing to random motion of stars.
Fig. 4 also shows that stars with positive  $\Delta Y$
are more likely to have positive $v_{\rm x}$,
which means that the simulated structure
has internal rotation
with the direction consistent with that of the Galactic disk.
The velocity gradient of  $v_{\rm x}$  for $|\Delta Y| \le 400$ pc
is 21.1 km s$^{-1}$ kpc$^{-1}$, which is comparable to
the angular velocity of $7-24$  km s$^{-1}$ kpc$^{-1}$ derived from
a kinematical model based on observational  data
sets of stars in the GB (Lindblad 2000).

\begin{figure}
\psfig{file=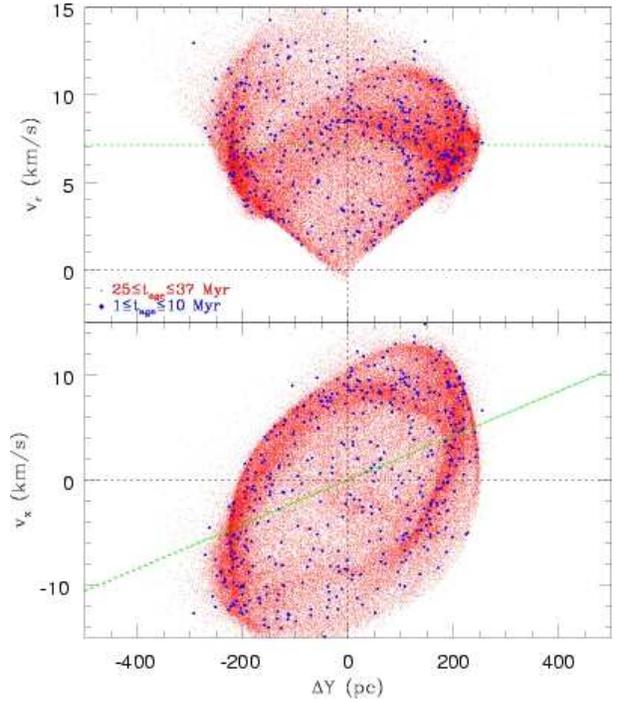,width=8.0cm}
\caption{
Kinematics of the simulated stellar structure.
The radial velocities ($v_{\rm r}$) and $X$-components of velocities
($v_{\rm x}$) for new stars
with 25 Myr $\le t_{\rm age}\le$ 37 Myr (smaller red dots)
and with 1 Myr $\le t_{\rm age}\le$ 10 Myr (bigger blue dots) are
plotted
against $\Delta Y$ in upper and lower frames, respectively.
Here $v_{\rm r}$ and $v_{\rm x}$ are velocity components
with respect to those of the mass center of the stellar structure.
The dotted green lines in the upper and lower panels represent the
mean value of $v_{\rm r}$ (= 7.1 km s$^{-1}$)
for all stellar particles and rigid rotation with
an angular velocity of 21.1 km s$^{-1}$ kpc$^{-1}$, respectively.
}
\label{Figure. 4}
\end{figure}

\section{Discussion and conclusions}

By using the standard formula for the timescale of collision/merging,
the time scale of a cloud-DMC collision event ($t_{\rm m}$)
can be estimated
as follows (Makino \& Hut 1997);
\begin{equation}
t_{\rm m}=\frac{ 1 } {n_{\rm h}\sigma v},
\end{equation}
where $n_{\rm h}$, $\sigma$, and $v$
are the mean number density of the DMCs within
the Galaxy halo,
the geometrical cross section of gas clouds
and a relative velocity between a DMC  and a cloud.
We here estimate $n_{\rm h}$ for the central 50 kpc
of the Galaxy (corresponding to the pericenter of the LMC orbit)
and assume that
$v$ is the same as one-dimensional
velocity dispersion ($=v_{\rm c}/\sqrt(2)$,
where $v_{\rm c}$ is the circular velocity thus 220 km s$^{-1}$)  of the
Galaxy
halo.
If we use the results of
the latest cosmological simulations,
we can estimate  the total number of DMCs ($N_{\rm h}$)
with masses larger than $\sim 10^7 {\rm M}_{\odot}$
within $50$ kpc (Diemand et al. 2008).
The total number of subhalos in the latest high-resolution
cosmological simulation (Springel et al. 2008a) is  by a factor
of $\sim 8$ larger than that of Via Lactea II (Diemand et al. 2008).
Considering this resolution effect,
we adopt $N_{\rm h}=100$ rather than $ N \approx 20$ (Diemand et al.
2008)
in the estimation
of $t_{\rm m}$.

We here estimate $t_{\rm m}$ for a DMC to collide with one
of many GMCs (not with a specific GMC).
Therefore  $\sigma$ should be $N_{\rm cl} \times \pi \times R_{\rm
cl}^2$,
where  $N_{\rm cl}$ and  $R_{\rm cl}$ are the total number of
the massive GMCs in the Galaxy and their sizes, respectively.
Given that the Galaxy has $\sim 4000$
massive GMCs (Solomon et al. 1987)  with masses more than $10^5 {\rm
M}_{\odot}$,
we can derive $t_{\rm m}$ for $N_{\rm cl}=4000$ as follows;
\begin{equation}
t_{\rm m}= 0.26
{ ( \frac{ N_{\rm h} } {100} )  }^{-1}
{ ( \frac{ R_{\rm cl} } {100 {\rm pc}} )  }^{-2}
{ ( \frac{ v } {156  {\rm km s^{-1} } } )  }^{-1}
{\rm Gyr}
\end{equation}
This suggests that the cloud-DMC  collision is not  rare
and thus that the dark impact can inevitably influence evolution
of the Galactic disk to some extent. The collisions between DMCs with
lower masses ($M_{\rm dm} \sim 10^6 {\rm M}_{\odot}$)
and less massive GMCs (with masses less than $10^5 {\rm M}_{\odot}$)
can be much more frequent than
those described above, because a much larger number of these less
massive DMCs
are predicted to exist in the Galactic halo (Diemand et al. 2008).
These collisions between less massive DMCs and GMCs can form
significantly smaller (thus less clearly visible)
stellar structures in comparison with
the GB.

The present study has first demonstrated that the dark impact can
dramatically change spatial distributions of forming stars
and star formation
histories within giant molecular clouds. This new mode of star formation
is in a striking contrast to those resulting from  small-scale
turbulence
within clouds (e.g.,  Larson 1981)
and from global galaxy-scale dynamical instability (Elmegreen 1995)
Future observational studies of stars in the GB  by GAIA (Perryman et
al. 2001)
will determine the three-dimensional structural and kinematical
properties
of the GB stars and
the individual stellar ages
and thus enable us to compare the results with those of the present
study.  This comparison will determine whether and when
the majority of stars in the GB  were formed from a single collisional
event of a DMC in a more quantitative way.
The GB is not just a magnificent  stripe of stars in the night sky
but also a fossil record containing profound physical meanings
of a dramatic event which would have happened
in the solar neighborhood.

\section{Acknowledgment}
I am   grateful to the referee Fernando Comeron for valuable comments,
which contribute to improve the present paper.
KB acknowledges the financial support of the Australian Research
Council
throughout the course of this work.
The numerical simulations reported here were carried out on GRAPE
systems at University of New South Wales
and those kindly made available by the Center for Computational
Astrophysics (CfCA)
at National Astronomical Observatory of Japan (NAOJ).

\end{document}